\documentclass[journal,twocolumn]{IEEEtran}

\def\BibTeX{{\rm B\kern-.05em{\sc i\kern-.025em b}\kern-.08em
    T\kern-.1667em\lower.7ex\hbox{E}\kern-.125emX}}
\usepackage{xcolor}
\usepackage{mathtools}
\usepackage{epsfig,makeidx,color,subfigure}
\usepackage{amsmath,amssymb,bbm,enumitem}
\usepackage{cite,graphicx,lipsum}
\usepackage[switch,pagewise]{lineno}
\usepackage{hyperref}
\hypersetup{
        colorlinks = true,
        citecolor=blue,
}
\def\rH{{\rm H}}
\def\rT{{\rm T}}

\def\uC{{\mathbb C}}
\def\uE{{\mathbb E}}

\DeclareMathOperator*{\argmin}{\arg\!\min}

\newtheorem{mytheorem}{\bf Theorem} 
\newtheorem{mylemma}{\bf Lemma} 
\newtheorem{myexample}{\it Example} 

\def\be{ \begin{equation} }
\def\ee{ \end{equation} }
\def\bea{ \begin{eqnarray} }
\def\eea{ \end{eqnarray} }

\def\bx{{\bf x}}
\def\by{{\bf y}}

\def\ba{{\bf a}}

\def\bn{{\bf n}}

\def\bz{{\bf z}}

\def\bA{{\bf A}}

\def\bF{{\bf F}}

\def\bI{{\bf I}}

\def\bR{{\bf R}}

\def\b0{{\bf 0}}

\def\cC{{\cal C}}

\def\cN{{\cal N}}

\def\sMSE{{\sf MSE}}
\def\sSNR{{\sf SNR}}

\ifCLASSOPTIONonecolumn
  \newcommand{\figwidth}{0.50\columnwidth}
  
\else
  \newcommand{\figwidth}{0.85\columnwidth}
  
\fi

\ifCLASSOPTIONonecolumn
\fi

\begin{document}

\title{Wireless Distributed Matrix-Vector Multiplication using Over-the-Air Computation and Analog 
Coding}

\author{
Jinho Choi\thanks{J. Choi is with School of Information Technology, 
Deakin University,
Email: jinho.choi@deakin.edu.au}
}


\maketitle
\begin{abstract}
In this paper, we propose an over-the-air (OTA)-based approach for distributed matrix-vector multiplications in the context of distributed machine learning (DML). 
Thanks to OTA computation, the column-wise partitioning of a large matrix enables efficient workload distribution among workers (i.e., local computing nodes) based on their computing capabilities. In addition, without requiring additional bandwidth, it allows the system to remain scalable even as the number of workers increases to mitigate the impact of slow workers, known as stragglers. However, despite the improvements, there are still instances where some workers experience deep fading and become stragglers, preventing them from transmitting their results. By analyzing the mean squared error (MSE), we demonstrate that incorporating more workers in the OTA-based approach leads to MSE reduction without the need for additional radio resources. Furthermore, we introduce an analog coding scheme to further enhance the performance and compare it with conventional coded multiplication (CM) schemes. Through simulations, it is shown that the OTA-based approach achieves comparable performance to CM schemes while potentially requiring fewer radio resources.
\end{abstract}

\begin{IEEEkeywords}
Distributed Machine Learning;  Over-the-Air Computation;
Analog Channel Coding
\end{IEEEkeywords}

\section{Introduction}

\subsection{Background}

Distributed machine learning (DML) has become an increasingly important research area due to the explosive growth of data and the need for scalable and efficient machine learning algorithms \cite{Verbraeken20} \cite{Wang22}. 
Federated learning \cite{Konecn2015FederatedOD} \cite{Li20}, as an example of DML, combines the power of collaborative and privacy-preserving training with decentralized computation. It has emerged as a transformative approach for training models directly on decentralized devices (called workers), such as mobile phones or edge devices, without the need to transfer sensitive data to a central server. This decentralized computation aspect not only addresses privacy concerns but also harnesses the vast amounts of data available on these devices for model training. By aggregating local model updates from multiple devices, federated learning enables collaborative learning while safeguarding the privacy and security of individual data sources. 

In federated learning, when the workers involved are mobile devices, wireless connectivity is utilized, which introduces the challenge of limited bandwidth \cite{Konecny16}. This limitation poses a bottleneck in the communication process and necessitates efficient compression techniques for transmitting the local gradient vectors which are crucial for model updates and aggregation in the federated learning framework, by mobile devices \cite{Alistarh17} \cite{Gandikota21a}. 

In aggregation, to harness the benefits of wireless communication, the concept of over-the-air (OTA) computation \cite{Nazer07} can be explored. In OTA computation, each mobile device does not require a dedicated channel to transmit its local gradient vector. Instead, all mobile devices simultaneously send their gradient vectors through a shared channel, leveraging the superposition nature of radio frequency (RF) signals. The server can receive and aggregate these transmitted vectors \cite{Yang20} \cite{Mohammadi20}, addressing the scalability issue associated with limited bandwidth. 
This approach enables more efficient and scalable communication protocols in large-scale distributed machine learning systems. 
In \cite{Choi22}, a random access scheme used in machine-type communication (MTC) \cite{Wunder15} \cite{Choi22_WC} is used to perform OTA computation with quantization.
The resulting scheme holds great potential for seamless integration into Internet-of-Things (IoT) devices with MTC capability. 

In addition to federated learning, another notable scenario in the context of DML involves performing large-scale matrix-vector multiplication using a number of workers \cite{Bertsekas2015parallel}. 
In this case, the objective is to distribute the matrix-vector multiplication task among a large pool of workers to expedite the computation process and achieve high-performance results. Each worker is responsible for performing a specific portion of the computation, typically involving a subset of rows 
of the matrix. By leveraging parallel processing and task distribution, this distributed approach enables significant acceleration of the computation compared to traditional centralized methods.

To ensure efficient coordination and synchronization among the workers, distributed computing frameworks and communication protocols are employed. These frameworks facilitate the partitioning of the matrix and vector across the workers, orchestrate the computation process, and aggregate the intermediate results to obtain the final output. Efficient load balancing mechanisms are crucial to evenly distribute the workload among the workers and minimize computation disparities, ensuring optimal resource utilization.

One of the key challenges in this context is mitigating the impact of stragglers—workers with slower processing capabilities or network connectivity issues \cite{Dean13}. Stragglers can introduce computational latency and hinder the overall performance of the distributed system.

In this paper, we extend the notion of OTA computation to perform large-scale matrix-vector multiplication using multiple workers in DML. By leveraging OTA computation, we introduce a novel approach to efficiently distribute the computational task of matrix-vector multiplication across a network of workers. Thanks to OTA computation, we exploit the superposition nature of wireless communication to enable the simultaneous transmission and aggregation of partial results from workers through column-wise partitioning, effectively mitigating the impact of stragglers by allocating tasks based on workers' capabilities. This OTA-based approach holds significant potential for enhancing the scalability and efficiency of large-scale matrix-vector multiplication in DML settings. Note that although OTA computation has been explored in the context of federated learning \cite{Yang20}, its application to distributed matrix-vector multiplication has not been addressed in the existing literature. Our work represents an application of OTA computation specifically to distributed matrix-vector multiplication, extending its prior utilization in federated learning scenarios

Throughout this paper, we aim to demonstrate the advantages of leveraging wireless communication and OTA computation in improving computation performance and resource utilization in DML scenarios. As a result, the primary performance metric is not the computation delay (due to stragglers), as stragglers can handle lighter computing tasks through flexible task allocation. Consequently, computation accuracy becomes a crucial performance measure, and our performance analysis is geared towards understanding computation accuracy under wireless channels.


\subsection{Existing Approaches}

In DML, certain workers can be stragglers due to \emph{i)} slow computing; or \emph{ii)} transmission errors, and the presence of stragglers can significantly impact its overall performance and efficiency. To mitigate stragglers, coding has been considered in 
\cite{Tandon17} \cite{Li17} \cite{Lee18}. 

In particular, the approach in \cite{Lee18} is considered for large-scale matrix-vector multiplications in DML. The overall completion time of a distributed computation depends on the performance of the slowest worker involved in the task. In scenarios where a fixed completion time is required, workers who are unable to complete their computations within the given time frame will not contribute their outputs. To mitigate this problem, the notion of coding can be used with more workers. 
For example, consider  a matrix-vector multiplication, $\by = \bA \bx$, where the sizes of $\bA$ and $\bx$ are $K \times L$ and $L \times 1$, respectively. If there are $K$ workers, each worker can perform $\ba_k^\rT \bx$, where $\ba_k^\rT$ represents the $k$th row of $\bA$. As mentioned earlier, the overall completion time to obtain $\by$
depends on that of the slowest worker, and with a fixed completion time, some elements of $\by$ assigned to stragglers might be unavailable. To mitigate this problem, there can be additional workers. To be specific, let $K = 2$ and assume that there is one additional worker who is to compute $(\ba_1^\rT + \ba^\rT_2) \bx$. We can see that as long as there are any two workers completing tasks within a given completion time, the server can obtain $\by = [y_1 \ y_2]^\rT$  (e.g., even if worker 2 cannot send $y_2 = \ba_2^\rT \bx$, the server can still find it from the outputs of workers 1 and 3 as $y_3 - y_1= (\ba_1^\rT + \ba^\rT_2) \bx - \ba_1^\rT \bx= y_2$). This can be seen as a form of channel coding, where the presence of a third worker is equivalent to introducing a redundant bit. 
In particular, in the context of channel coding, the presence of stragglers can be analogized to the occurrence of erased packets in a communication channel.  By leveraging erasure coding techniques, such as maximum distance separable (MDS) codes, the adverse effects of stragglers can be mitigated.

In \cite{Reisizadeh19}, heterogeneous workers in terms of computing power are considered, which can be seen as a generalization of the work in \cite{Lee18} where all workers have the same completion time distribution.
In \cite{Yu20}, an optimal code design is studied in terms of the recovery threshold, which is the minimum number of workers required for the server to successfully recover the desired output.
In \cite{Mallick22}, the use of rateless codes \cite{Luby02} is considered.
A survey of coding techniques to mitigate stragglers can be found in
\cite{Xiao22}. For convenience, the approaches based on (digital channel) coding (e.g., the approaches in \cite{Lee18} \cite{Mallick22}) to mitigate stragglers are referred to as coded multiplication (CM) schemes in this paper. 

When matrices are sparse, coding schemes, such as MDS codes, typically generate dense linear combinations of submatrices, leading to the loss of inherent sparsity and longer worker computation times compared to scenarios where sparsity is effectively exploited. To address this issue, coding schemes that specifically aim to exploit sparsity in order to reduce computational complexity are investigated in \cite{Wang18e} \cite{Das18} \cite{Das21}. Additionally, the work presented in \cite{Xhemrishi22} also takes privacy considerations into account within the context of coding schemes.

It is noteworthy that the majority of existing approaches for distributed matrix-vector multiplication do not incorporate considerations for transmissions over wireless channels. While these approaches address various challenges such as heterogeneous workers \cite{Reisizadeh19}, sparse matrices \cite{Das21}, or privacy issues \cite{Xhemrishi22}, they typically overlook the nuances introduced by wireless connections. Given the increasing prevalence of a large number of mobile workers, it becomes imperative to account for wireless connectivity. In this paper, our emphasis is on the challenges posed by distributed matrix-vector multiplication with mobile workers connected through wireless channels, setting our work apart from existing approaches.


\subsection{Main Contributions}

In this paper, we consider a different approach based on OTA computation in performing distributed matrix-vector multiplication when workers are connected through wireless channels. In most CM schemes (e.g., \cite{Lee18}), since each worker is responsible to compute some elements of $\by$ or their linear combinations, the computational complexity is proportional a multiple of $L$, i.e., the size of $\bx$. 
In other words, the computational complexity for each worker is directly influenced by the value of $L$, and as $L$ increases, the likelihood of workers experiencing straggling issues also rises. To overcome this challenge, we propose a solution that involves partitioning the large matrix $\bA$ into multiple smaller submatrices. This partitioning allows each worker to perform the multiplication of a smaller submatrix with a corresponding subvector. By distributing the workload in this manner, even in scenarios where $K$ is significantly large, the computational load can be properly allocated based on each worker's computing power, mitigating the occurrence of stragglers, particularly in heterogeneous scenarios. That is, this approach enables more efficient and balanced processing across the workers, with a smaller granularity of computational complexity, ultimately mitigating the impact of stragglers and enhancing the overall performance of the system. However, if each worker sends it result through a dedicated orthogonal channel, this approach needs more bandwidth. To avoid this problem, OTA computation can be used where multiple workers can transmit their results simultaneously through a shared channel. Then, taking advantage of the superposition nature of wireless communication, the server can then receive and aggregate these transmitted results, enabling a scalable and communication-efficient solution that mitigates straggler effects and enhances overall system performance. In summary, the first contribution is as follows.
\begin{itemize}
\item[C1)] We propose an approach based on OTA computation that allows to lower the granularity of computation so that each worker can have a task within their computational capability without increasing the required bandwidth.
\end{itemize}
While the proposed OTA-based approach can remove the problem caused by stragglers, there exists another critical problem due to wireless channels. Some workers' channels may experience deep fading \cite{TseBook05} and their results cannot be reliably transmitted to the server. Thus, it is important to understand the performance of the proposed approach in terms of mean squared error (MSE). 
Thus, the second contribution is as follows.
\begin{itemize}
\item[C2)] The MSE of the proposed approach is analyzed with a maximum transmit power and a bound as a closed-form expression is obtained.
\end{itemize}

The notion of channel coding can be considered to  improve the performance at the cost of bandwidth \cite{WickerBook}. Since the OTA-based approach is considered, to fully exploit the benefits of OTA computation, it becomes necessary to use analog coding techniques. Analog coding involves representing information using continuous, real-valued signals, as opposed to discrete binary values used in digital coding. The third contribution of the paper is as follows.
\begin{itemize}
\item[C3)] An analog coding approach is proposed for the OTA-based approach and its performance is evaluated through analysis using a random coding setup.
\end{itemize}


The rest of the paper is organized as follows.
In Section~\ref{S:SM}, the proposed approach based on OTA computation for distributed matrix-vector multiplication is presented together with the system model. The MSE analysis and a coding approach are discussed in Section~\ref{S:Coding} to see the impact of stragglers on the performance in terms of MSE. In Section~\ref{S:Anal}, the performance is analyzed under fading channels. Simulation results are presented in Section~\ref{S:Sim}, and the paper is concluded with some remarks in Section~\ref{S:Con}.

\subsubsection*{Notation}
Matrices and vectors are denoted by upper- and lower-case
boldface letters, respectively.
The superscripts $\rT$ and $\rH$
denote the transpose and complex conjugate, respectively.
$\bI_n$ represents an $n \times n$ identity matrix.
$\uE[\cdot]$
and ${\rm Var}(\cdot)$
denote the statistical expectation and variance, respectively.
$\cC \cN(\ba, \bR)$
represents the distribution of
circularly symmetric complex Gaussian (CSCG)
random vectors with mean vector $\ba$ and
covariance matrix $\bR$.

\section{System Model}  \label{S:SM}

\subsection{Column-wise Partitioning for OTA Computation}

\begin{figure}[!t]\vspace{-1em}
\begin{center} \subfigure[Broadcasting phase]{\label{fig 1 ax}\includegraphics[width=0.45\textwidth]{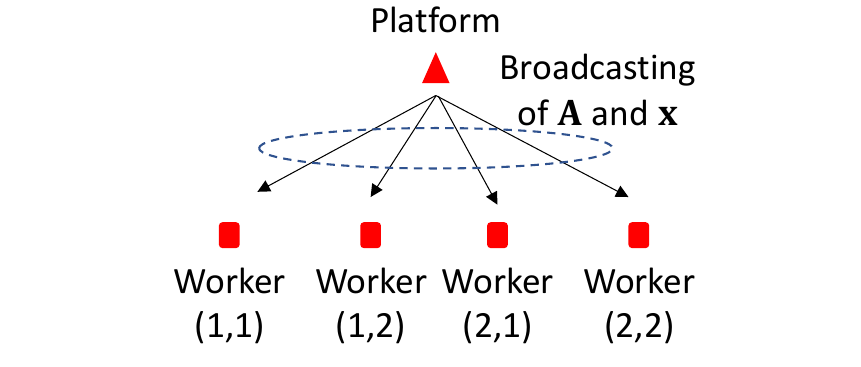}}
\subfigure[Superposition phase]{\label{fig 1 bx}\includegraphics[width=0.45\textwidth]{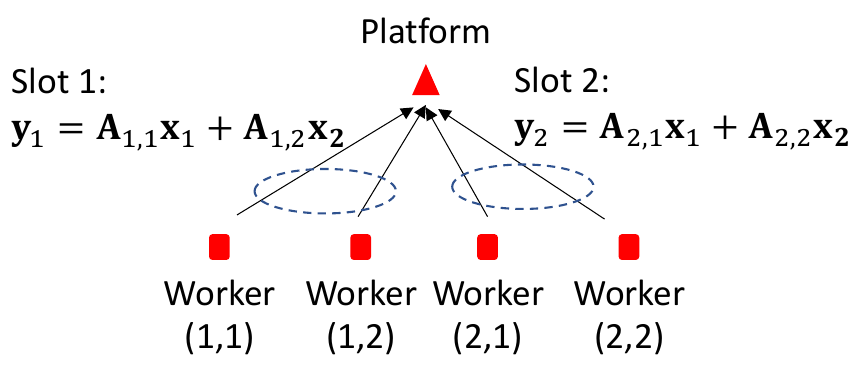}} \vspace{-1em}
\end{center}
\caption{An example of two-phase communication between a platform and multiple workers for distributed machine learning.}\label{Fig:1}
\end{figure}

Suppose that the platform (or server) needs to compute a matrix-vector multiplication, $\by = \bA \bx$, and there are 4 workers that are connected to the platform through wireless channels.
Then, the matrix is partitioned into 4 sub-matrices as follows:
$$
\bA = \left[
\begin{array}{cc} 
\bA_{1,1} & \bA_{1,2} \cr 
\bA_{2,1} & \bA_{2,2} \cr 
\end{array}
\right].
$$
Then, $\by = [\by_1^\rT \ \by_2^\rT]^\rT$ can be obtained by
\begin{align*}
\by_1 & = \bA_{1,1} \bx_1 + \bA_{1,2} \bx_2 \cr 
\by_2 & = \bA_{2,1} \bx_1 + \bA_{2,2} \bx_2,
\end{align*}
where $\bA_{k,l} \bx_l$ is computed at worker $(k,l)$, $k, l \in \{1,2\}$.
Here, for convenience, the indices of workers are represented by a tuple of two indices,
$k$ and $l$.
In each round, there are two phases for downlink and uplink, which are illustrated in Fig.~\ref{Fig:1}, as follows.
\begin{enumerate}
\item Downlink: This is the \emph{broadcasting phase} to send the necessary data to all workers as shown in Fig.~\ref{fig 1 ax}. Due to the broadcasting nature of wireless communications, when the platform transmits $\bA$ as well as $\bx$, all the workers can receive them.
\item Uplink: This is the \emph{superposition phase} for the workers to send their outcomes of tasks as shown in Fig.~\ref{fig 1 bx}. This phase is divided into $K = 2$ subslots. In slot $k$, the two workers to perform tasks to compute $\by_{k,l} = \bA_{k,l} \bx_l$, 
$l= 1,2$, transmit their results\footnote{Throughout the paper, we assume that each element of $\by_{k,l}$ is transmitted at the Nyquist rate (i.e., the inverse of the system bandwidth). As a result, one element of $\by_{k,l}$ is to be transmitted within one symbol duration.}. 
Then, thanks to the superposition nature of wireless communications, the platform can receive $\sum_{l=1}^2 \bA_{k,l} \bx_l$ during slot $k \in \{1,2\}$ through the OTA computation. 
\end{enumerate}
For convenience, this approach is referred to as the OTA-based approach.

In the OTA-based approach, the salient feature is that column-wise partitioning can be employed to divide the computational task among multiple workers without requiring  additional bandwidth. To see this clearly, consider 3 different cases as follows:
\begin{itemize}
\item[1)] No column-wise partitioning is considered, where
worker $k \in \{ 1,\ldots, 4 \}$ computes $\ba_k^\rT \bx$.
\item[2)] Column-wise partitioning is considered, where
worker $(k,l)$ computes $\bA_{k,l} \bx_l$ without OTA computation.
\item[3)] The same as Case 2 except OTA computation, i.e., OTA-based approach.
\end{itemize}
In Case 1,
assuming that a unit uplink channel is required to send $y_k$, there should be 4 unit uplink channels or 4-symbol duration. In Case 2, workers $(k,1)$ and $(k,2)$ needs to compute $\bA_{k,1} \bx_1$ and $\bA_{k,2} \bx_2$, respectively, and transmit the results through separate channels without OTA computation. Thus, there should be two times more uplink channels required, i.e., 8 unit uplink channels. On the other hand, in Case 3, thanks to OTA computation, 4 unit uplink channels are required, which is the same as Case 1. 



However, when no column-wise partitioning is considered, the granularity of the computational complexity of the task per each worker is limited. That is, the number of multiplications to perform $\ba_k^\rT \bx$, which is the length of $\bx$, can be considered to be a minimum computational complexity required.  As a result, when the length of $\bx$ increases, the potential for computational latency introduced by stragglers, i.e., workers with lower computing power than the required computational complexity, also increases.

On the other hand, in the OTA-based approach, the workers can have different computational complexity of tasks. Denote by $C(k,l)$ the computing power of worker $(k,l)$. Since the number of multiplications for  worker $(k,l)$ is the same as the number of the elements of $\bA_{k,l}$, denoted by $|\bA_{k,l}|$, slower workers can have the tasks with smaller $|\bA_{k,l}|$'s through load balancing. That is, the task allocation can be carried out to keep the variation of the ratio $C(k,l)/|\bA_{k,l}|$ minimum as the sizes of $\bA_{k,l}$ are controllable, which makes the OTA-based approach flexible to mitigate stragglers without resorting to  more workers.
However, due to channel impairments (especially fading), it would also be necessary to consider a coding scheme, which will be discussed in Section~\ref{S:Coding}.

\begin{figure}[thb]
\begin{center}
\includegraphics[width=\figwidth]{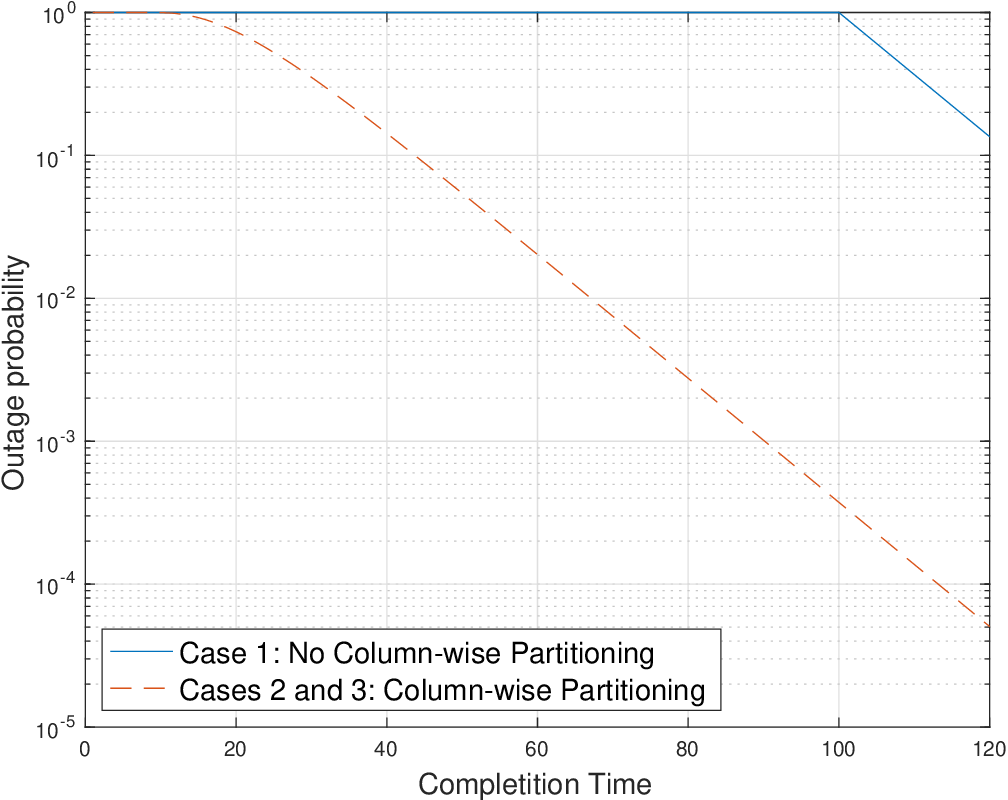} 
\end{center}
\caption{Outage probabilities for the 3 different cases.}
        \label{Fig:exam1}
\end{figure}

\begin{myexample}
In this example, we consider the above 3 different cases with $\bA$ of a size of $K \times L$ and $K = 4$ workers with $L = 100$. The task completion time of worker $k$ can be found as
$$
T_k = \frac{G_k}{c_k} + Z_k,
$$
where $G_k$ is the number of central processing unit (CPU) cycles required to perform a given task, $c_k$ is the clock speed of CPU (in the number of CPU cycles per second), and $Z_k$ is the additional time including initial setup time and processing delay for reception and transmission. 
In addition, assume that $c_1 = 1, c_2 = c_3 = 10$, and $c_4 = 40$, while $Z_k \sim {\rm Exp}(\mu)$, where $\uE[Z_k] = \frac{1}{\mu} = 10$. 
For Case 1, we have $G_k = \tau L$ (e.g., each worker is to compute $\ba_k^\rT \bx$), where $\tau > 0$ is constant (which is assumed to be unity for convenience). 


For Cases 2 and 3, with column-wise partitioning, let the sizes of $\bA_{1,1}$ (for worker 1), $\bA_{1,2}$ (for worker 2), $\bA_{2,1}$ (for worker 3), and $\bA_{2,2}$ (for worker 4) be $1 \times L_1$, $1\times L_2$, $3 \times L_1$, and $3 \times L_2$, respectively, where $L_1 + L_2 = L = 100$. For a load balancing, we can choose $L_1 = 9$ so that $\max_k \frac{L_k}{c_k}$ is minimized. We consider the outage probability that is given by $\Pr(\max_k T_k > d)$, where $d$ represents the overall completion time, and show the results for the 3 different cases in Fig.~\ref{Fig:exam1}. In Case 1, the overall completion time is mainly decided by worker 1 (with $c_1 = 1$) who has the slowest CPU, since the computation loads are the same for all the workers. However, in Cases 2 and 3, thanks to column-wise partitioning, the computation load can be assigned according to workers' computing powers with a finer granularity for load balancing and a lower outage probability can be achieved.
\end{myexample}

Throughout the paper, thanks to column-wise partitioning, it is assumed that an ideal load balancing is carried out according to the workers' computing powers in the OTA-based approach. Thus, 
no stragglers due to limited computing power are considered.

\subsection{System Model of Wireless DML}

Although ideal load balancing is assumed, there can be stragglers due to transmission errors. In this subsection, we present the system model with fading channels for wireless DML to see the impact of stragglers due to transmission errors.

We assume that there are $J = KL$ workers and a tuple of $k$ and $l$,
$k\in \{1,\ldots, K\}$ and $l \in \{1,\ldots, L\}$ is associated with a worker, i.e., worker $(k,l)$. Furthermore, a time-slotted system is assumed, where $K$ time slots are allocated for the superposition phase. Workers $(k,l)$, $l \in \{1,\ldots, L\}$, belong to group $k$ that transmit their outcomes, i.e., $\bA_{k,l} \bx_l$, in slot $k$.
Let the size of $\bA_{k,l}$ be $M_k \times Q_l$. Thus, the size of $\bA$ becomes $M \times Q$, where $M = \sum_{k=1}^K M_k$ and $Q = \sum_{l=1}^L Q_l$. As a result, the length of time slot $k$ is $M_k$ in unit time or $M_k$-symbol duration and there are $L$ workers in each group $k$, $k \in \{1,\ldots, K\}$.


It is assumed that the communication between the platform and workers are based on the time division duplexing (TDD) mode, and 
the platform transmits a pilot signal in the broadcasting phase together with the data, $(\bA, \bx)$. 
The channel coefficient from the platform to worker $(k,l)$ is denoted by $h_{k,l}$, which remains unchanged during each communication round consisting of the broadcasting and superposition phases. 
Thus, the received signal at the platform in the $k$th slot during the superposition phase is given by
\be 
\bz_k = \sum_{l=1}^L h_{k,l} g_{k,l} \by_{k,l} + \bn_k \in \uC^{M_k},
    \label{EQ:zk}
\ee 
where $g_{k,l}$ is the (channel) compensation coefficient of worker $(k,l)$ for OTA computation, $\by_{k,l} = \bA_{k,l} \bx_l$ is the output of worker $(k,l)$, and $\bn_k \sim \cC \cN (0, N_0 \bI)$ is the background noise. 

Due to the channel reciprocity of TDD, each worker knows the channel coefficient, $h_{k,l}$. Thus, the compensation coefficient  of each worker can be decided to meet the following requirement:
\be 
h_{k,l} g_{k,l} \ge \sqrt{P_{\rm rx}},
    \label{EQ:hg}
\ee 
where $P_{\rm rx}$ represents the desired received signal power at the platform. For simplicity, we consider the equality in the above equation, which also establishes the minimum transmit power satisfying the requirement.
Substituting \eqref{EQ:hg} into \eqref{EQ:zk}, we have
\be 
\bz_k = \sqrt{P_{\rm rx}} \by_k + \bn_{k}, \ k = 1, \ldots, K,
    \label{EQ:zk2}
\ee 
where $\by_k = \sum_{l=1}^L \bA_{k,l} \bx_{l}$, which can be seen as an (scaled) estimate of $\by_k$.

A worker can be a straggler due to transmission errors
in the OTA-based approach. Since the superposition phase relies on analog communications, a reliable linear power amplifier has to be used for each worker, which may limit the peak transmit power. This means that there might be workers who may not be able to compensate the channel distortion as in \eqref{EQ:hg}  when $|h_{k,l}|$ is small (e.g., due to deep fading) and becomes a straggler, although ideal loading balancing is used.

Note that throughout the paper, we only consider the impact of stragglers during the superposition phase on the performance. In fact, since TDD is assumed, stragglers (i.e., workers under deep fading) likely fail to receive the data from the platform during the broadcasting phase. However, to simplify analysis, we assume that all the workers receive the data (i.e., $\bA$ and $\bx$) without errors and study the performance degradation in the superposition phase.


\section{Mitigating Stragglers}   \label{S:Coding}

In this section, we will discuss the impact of channel impairments on the performance in the MSE and show that more workers can help improve the performance and also propose a coding scheme.

\subsection{MSE Analysis}

As mentioned earlier, the peak transmit is usually limited by the maximum transmit power. Thus, 
letting
\be
\psi_{k,l} = \max_i |[\by_{k,l}]_i|^2,
    \label{EQ:psi}
\ee
where $[\bx]_i$ represents the $i$th element of $\bx$, the peak transmit power has to be limited to satisfy the following inequality:
\be 
P_{k,l}^\ast = |g_{k,l}|^2 \psi_{k,l} \le \bar P_{k,l} ,
    \label{EQ:pkl}
\ee 
where $\bar P_{k,l}$ represents the maximum transmit power of worker $(k,l)$. 
Due to either a low channel gain $|h_{k,l}|$ or a large maximum amplitude of the outcome  $\psi_{k,l}$, the required power $P_{k,l}^\ast$ of a worker can be higher than $\bar P_{k,l}$, meaning that
some workers may not be able to determine $g_{k,l}$ according to \eqref{EQ:hg}.
In this case,  an optimization approach can be used to determine $g_{k,l}$ while taking into account the constraint specified in \eqref{EQ:pkl}.

In the superposition phase, the platform receives an estimate of $\by_k$, which is $\frac{1}{\sqrt{P_{\rm rx}}} \bz_k$ and the MSE of slot $k$, $k = 1,\ldots, K$, can be given by
\begin{align} 
\epsilon_k 
& = \uE \left[ ||\frac{1}{\sqrt{P_{\rm rx}}} \bz_k - \by_k ||^2 \right] \cr
& = \uE \left[
||\frac{1}{\sqrt{P_{\rm rx}}} \bz_k - \sum_{l=1}^L \by_{k,l} ||^2 \right] \cr 
& = \sum_l |1 - h_{k,l} g_{k,l}|^2 \uE[ ||\by_{k,l}||^2] +\frac{N_0}{P_{\rm rx}} M_k,
    \label{EQ:mse}
\end{align}
where the third equality is valid under the assumption that 
the $\by_{k,l}$'s have zero-mean and uncorrelated,
i.e., $\uE[\by_{k,l}] = \b0$ and $\uE[ \by_{k,l} \by_{k,l^\prime}^\rH] = \b0$ for $l \ne l^\prime$.
Then, to minimize the MSE, worker $(k,l)$ can decide her compensation coefficient by solving the following optimization problem:
\begin{align}
\hat g_{k,l} & = \argmin_{g_{k,l}} \epsilon_k      \cr 
\mbox{subject to} \ &  |g_{k,l}|^2 \psi_{k,l} \le \bar P_{k,l}.
    \label{EQ:OP}
\end{align} 
As the MSE in \eqref{EQ:mse} is a sum of individual MSE terms, worker $(k,l)$ can easily decide the optimal compensation coefficient that minimizes the MSE, i.e., $\hat g_{k,l}$, without knowing the others' coefficients as follows:
\begin{align}
\hat g_{k,l} = \frac{ h_{k,l}^\ast \uE[ ||\by_{k,l}||^2] }{
| h_{k,l}|^2 \uE[ ||\by_{k,l}||^2]
+ \lambda \psi_{k,l}},
    \label{EQ:hgkl}
\end{align}
where the superscript $\ast$ represents the complex conjugate and
$\lambda$ is the Lagrange multiplier that is to be decided to meet the constraint in \eqref{EQ:pkl}.

Note that if \eqref{EQ:pkl} is satisfied with $\lambda = 0$ at worker $(k,l)$, the corresponding MSE term in \eqref{EQ:mse} becomes 0.  
Otherwise, from \eqref{EQ:pkl}, we have 
$\lambda > 0$ if 
\be 
\frac{\psi_{k,l}}{|h_{k,l}|^2} > \bar P_{k,l}.
    \label{EQ:phP}
\ee 
In addition, when $\lambda > 0$,
the individual MSE is bounded as follows:
\begin{align}
\sMSE_{k,l} & = 
|1 - h_{k,l} \hat g_{k,l}|^2 \uE[ ||\by_{k,l}||^2] \cr
& = 
\frac{
(\lambda \psi_{k,l})^2 \uE[ ||\by_{k,l}||^2]
}
{ \left(
| h_{k,l}|^2 \uE[ ||\by_{k,l}||^2]
+ \lambda \psi_{k,l} \right)^2 } \cr 
& \le \uE[ ||\by_{k,l}||^2].
    \label{EQ:bMSE0}
\end{align}
We have an improve result in the following theorem.
\begin{mytheorem}
A tighter bound on the normalized individual MSE is given by
\be 
\frac{\sMSE_{k,l}}{\uE[ ||\by_{k,l}||^2]} 
\le
\min \left\{1, \frac{1}{4} \left(
\sqrt\frac{\psi_{k,l}}{|h_{k,l}|^2 \bar P_{k,l}} - 1 \right)^+ \right\},
     \label{EQ:bMSE1}
 \ee     
where $(x)^+ = \max\{0,x\}$.
\end{mytheorem}
\begin{IEEEproof}
Since \eqref{EQ:bMSE0} is shown, we only need to show 
\be 
\sMSE_{k,l} \le
\frac{1}{4} \left(
\sqrt\frac{\psi_{k,l}}{|h_{k,l}|^2 \bar P_{k,l}} - 1 \right)^+
 \uE[ ||\by_{k,l}||^2]. 
    \label{EQ:14}
\ee
According to \eqref{EQ:phP}, if $\frac{\psi_{k,l}}{|h_{k,l}|^2} \le \bar P_{k,l}$, $\lambda$ becomes 0 and $\sMSE_{k,l} = 0$, while
$\left(
\sqrt\frac{\psi_{k,l}}{|h_{k,l}|^2 \bar P_{k,l}} - 1 \right)^+ = 0$. Thus, we only need to consider the case that $\frac{\psi_{k,l}}{|h_{k,l}|^2} > \bar P_{k,l}$, where $\lambda > 0$, to derive \eqref{EQ:14}.

From the second equality in \eqref{EQ:bMSE0}, using the inequality of arithmetic and geometric means, it can be shown that
\begin{align}
\sMSE_{k,l}
& =\frac{\lambda \psi_{k,l}}{| h_{k,l}|^2} \frac{
\lambda \psi_{k,l} | h_{k,l}|^2 \uE[ ||\by_{k,l}||^2]
}
{ \left(
| h_{k,l}|^2 \uE[ ||\by_{k,l}||^2]
+ \lambda \psi_{k,l} )\right)^2 } \cr 
& \le \frac{1}{4} \frac{\lambda \psi_{k,l}}{| h_{k,l}|^2}.
    \label{EQ:uMSE0}
\end{align}
Since $\lambda > 0$, $\lambda$ is to satisfy the constraint with equality in \eqref{EQ:OP}. Thus, from \eqref{EQ:hgkl}, we have
\be 
\frac{|h_{k,l} |^2 (\uE[ ||\by_{k,l}||^2] )^2 \psi_{k,l}}{
(| h_{k,l}|^2 \uE[ ||\by_{k,l}||^2]
+ \lambda \psi_{k,l} )^2} = \bar P_{k,l}.
    \label{EQ:lam1}
\ee
By solving \eqref{EQ:lam1}, we have
\be
\frac{\lambda \psi_{k,l}}{|h_{k,l}|^2}
= \left(
\sqrt\frac{\psi_{k,l}}{|h_{k,l}|^2 \bar P_{k,l}} - 1 \right)
 \uE[ ||\by_{k,l}||^2],
    \label{EQ:lam2}
\ee 
where $\frac{\psi_{k,l}}{|h_{k,l}|^2 \bar P_{k,l}} > 1$ from \eqref{EQ:phP}.
Substituting \eqref{EQ:lam2} into \eqref{EQ:uMSE0}, we can show \eqref{EQ:14}, which completes the proof.
\end{IEEEproof}

\begin{figure}[thb]
\begin{center}
\includegraphics[width=\figwidth]{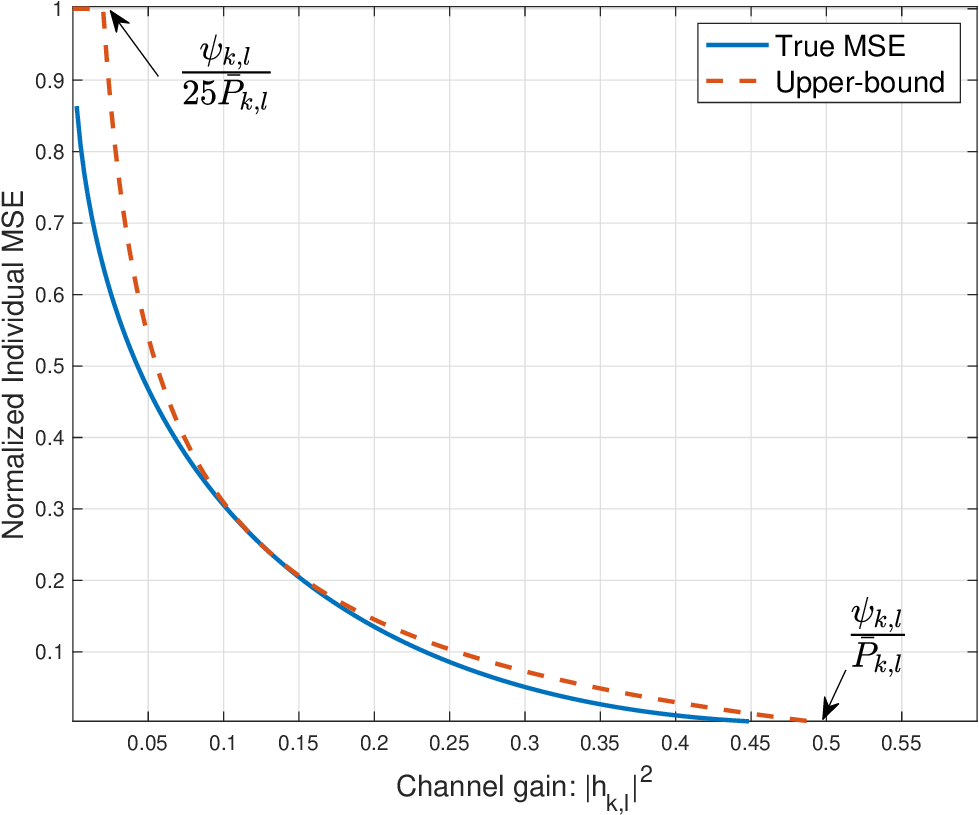} 
\end{center}
\caption{Normalized individual MSE with the upper-bound in \eqref{EQ:bMSE1}.}
        \label{Fig:plt_MSE}
\end{figure}

In Fig.~\ref{Fig:plt_MSE}, a normalized individual MSE as a function of channel gain, $|h_{k,l}|^2$, is shown with its upper-bound derived in \eqref{EQ:bMSE1} when $\psi_{k,l} = 1$, $\bar P_{k,l} = 2$, and $\uE[ ||\by_{k,l}||^2] = 5$. In \eqref{EQ:bMSE1}, it can be readily shown that the upper-bound on the normalized individual MSE becomes 1 when 
$\frac{\psi_{k,l}}{25 \bar P_{k,l}} = |h_{k,l}|^2$. From Fig.~\ref{Fig:plt_MSE}, we can see that the upper-bound is tight.

From \eqref{EQ:bMSE1}, it can be observed that the MSE has the potential to decrease with an increase in the maximum transmit power. Additionally, a decrease in $\psi_{k,l}$ can also contribute to the reduction of MSE. As shown in \eqref{EQ:psi},  $\psi_{k,l}$ represents the maximum of the squared elements of the vector $\by_{k,l}$. To decrease $\psi_{k,l}$, which in turn decreases the MSE, the length of $\by_{k,l}$, i.e., $M_k$, can be reduced. This reduction in $M_k$ can be achieved by increasing the number of groups, $K$, which results in the increase of the total number of workers, $J = KL$. The related simulation results are shown in Fig.~\ref{Fig:plt3} in Section~\ref{S:Sim}.

There is another way to decrease $\psi_{k,l}$ (or the MSE). As $\by_{k,l} = \bA_{k,l} \bx_l$, each element is a sum of $Q_l$ terms of $\bx$. Therefore, reducing $Q_l$ can lead to a decrease in $\psi_{k,l}$. However, the total number of columns in $\bA$,  $\sum_{l=1}^L Q_l$, remains fixed. To decrease $Q_l$, it is necessary to increase $L$, which in turn increases the total number of workers, $J = KL$. The related simulation results are shown in Fig.~\ref{Fig:plt5} in Section~\ref{S:Sim}.

\subsection{A Coding Scheme}

In the OTA-based approach, if each worker has a sufficient transmit power to meet \eqref{EQ:pkl}, the MSE only has a background noise term (i.e., it becomes $\frac{N_0}{P_{\rm rx}} M_k$ for each slot). 
However, the MSE increases if some workers cannot meet the power constraint in \eqref{EQ:pkl} due to \emph{i)} deep fading (i.e., small $|h_{k,l}|^2$); and/or \emph{ii)} a large peak transmit power (i.e., large $\psi_{k,l}$). To mitigate this problem, the notion of channel coding can be considered.

Consider  two matrices for encoding and decoding, denoted by $\bF$ and $\bF^\dagger$, respectively, which satisfy the following relationship:
\be 
\bI = \bF^\dagger \bF,  
    \label{EQ:FF}
\ee 
where $\bF \in \uC^{N \times M}$ and $\bF^\dagger \in \uC^{M \times N}$ with $N \ge M$. 
Then, for coding, $\tilde \bA = \bF \bA$, which is referred to as the coded data matrix, is to be sent to workers during the broadcasting phase. 
The signal received at the platform during the superposition phase has to be processed through decoding, i.e., the estimate of $\by$ becomes 
\be 
\hat \by = \frac{1}{\sqrt{P_{\rm rx}}} \bF^\dagger \bz,
    \label{EQ:hy}
\ee 
where $\bz = [\bz_1^\rT \ \ldots \ \bz_K^\rT]^\rT$. 
Due to \eqref{EQ:FF}, we can see that $\hat \by = \bA \bx + \frac{1}{\sqrt{P_{\rm rx}}}
\bF^\dagger \bn$ if all the workers can send their outcomes with satisfying \eqref{EQ:pkl}. Here, $\bn = [\bn_1^\rT \ \ldots \ \bn_K^\rT]^\rT$.
Note that The length of the superposition phase needs to be extended from $M$- to $N$-symbol duration, where 
$$
r = \frac{M}{N}
$$
can be seen as the code rate. Thus, coding requires more radio resources.

\begin{myexample}
Suppose that
$$
\bF^\dagger = \left[\frac{1}{2} \bI_M \ \frac{1}{2} \bI_M \right] \ \mbox{and} \ 
\bF = \left[ \begin{array}{c} \bI_M \cr \bI_M \cr 
\end{array} \right],
$$
where $N = 2M$. In this case, there should be twice as many workers  compared to the uncoded case (i.e., $\bF = \bI_M$).
In particular, two workers are assigned to the same task. The primary benefit of coding is that it can mitigate the degradation in MSE caused by deep fading, albeit at the expense of additional radio resources or workers. To see this, consider two workers with the task to compute $\bA \bx$ with assuming that $K = 1$ and $L = 1$. In this case, one worker sends her signal in slot 1 and the other sends in slot 2.
If the two workers can satisfy \eqref{EQ:pkl}, from \eqref{EQ:hy}, the platform can have the following estimate:
\begin{align*}
\hat \by & = \frac{1}{\sqrt{P_{\rm rx}}} \bF^\dagger 
[\bz_1^\rT \ \bz_2^\rT]^\rT 
= \frac{1}{2\sqrt{P_{\rm rx}}} ( \bz_1 + \bz_2)
\cr 
& = \by  + \frac{1}{2 \sqrt{P_{\rm rx}}} (\bn_1 + \bn_2),
\end{align*}
and the MSE becomes $\frac{N_0 M}{2 P_{\rm rx}}$.
On the other hand, if one of the workers (say worker 1) cannot meet \eqref{EQ:pkl}, the MSE becomes
\begin{align*}
\sMSE & = 
\uE \left[ ||\frac{1}{2 \sqrt{P_{\rm rx}}} (\bz_1 + \bz_2) - \by ||^2 \right] \cr
& = \frac{1}{4} |1 - h_{1} g_{1}|^2 \uE[ ||\by ||^2] +\frac{N_0 M}{2 P_{\rm rx}},
\end{align*}
which is a quarter of its value without coding. This example clearly demonstrates the advantage of coding (i.e., the decrease of MSE at the expense of workers and channel resources).
\end{myexample}

For analog coding, we may need to increase the number of workers as demonstrated earlier. For example, suppose that $M = 4$ and $Q = 8$. If each worker can compute a matrix-vector multiplication with a matrix of size $2 \times 2$, there should be $J = 8$ workers with $K = 2$ and $L = 4$. If analog coding is used with $N = 6$, there should be $J = 12$ workers as the size of coded data matrix, $\tilde \bA$, becomes $6 \times 8$. Thus, provided that the size of the data matrix remains the same for all workers, the proposed coding scheme requires an additional factor of $\frac{1}{r}$ in the number of workers, while the MSE can decrease.

There is however a problem when $M$ is large, which is a typical case in DML. The computational complexity for encoding, i.e., $\bF \bA$, and decoding, i.e., $\bF^\dagger \bz$, would be high and offsets the advantage of distributed computation. 
The number of multiplications for encoding, $\bF \bA$, is $O(MNL)$, while that for decoding, $\bF^\dagger \bz$, is $O(MN)$.
To avoid this problem, we may need to consider block-wise encoding and decoding with the encoding and decoding matrices for each slot, denoted by $\bF_k
\in \uC^{N_k \times M_k}$ and $\bF_k^\dagger \in \uC^{M_k \times N_k}$, respectively, where
it is assumed that
\be 
\bF_k^\dagger \bF_k = \bI_{M_k}, \ k = 1,\ldots, K.
\ee 
Thus, worker $(k,l)$ sends the outcome, $\tilde \bA_{k,l} \bx_l$, where $\tilde \bA_{k,l} = \bF_k \bA_{k,l} \in \uC^{N_k \times Q_l}$. 

Note that this block-wise approach is a special case, which can be readily shown by defining $\bF$ as
$$
\bF = \left[
\begin{array}{cccc}
\bF_1 & \b0 & \cdots & \b0 \cr 
\b0 & \bF_2 & \cdots & \b0 \cr 
\vdots & \vdots & \ddots & \vdots \cr 
\b0 & \b0  & \cdots & \bF_K\cr 
\end{array} \right],
$$
while $\bF^\dagger$ can be found similarly so that $\bF^\dagger \bF = \bI$.

To see the complexity of the block-wise approach, we can show that
\begin{align}
\tilde \bA_k & = \bF_k \bA_k \cr 
& = \bF_k [\bA_{k,1} \ \cdots \ \bA_{k,L}] \cr 
& = [\tilde \bA_{k,1} \ \cdots \ \tilde \bA_{k,L}].
\end{align}
Thus, for the workers in group $k$, the number of multiplications for encoding is $O(M_k N_k L)$, and the total number of multiplications becomes $O(\sum_k M_k N_k L)$. For example, if $M_k = \frac{M}{K}$ and $N_k = c M_k$, where $c$ is a constant, we have
\begin{align*}
\mbox{Complexity for encoding with} \ \bF \bA &= O( M^2 L) \cr 
\mbox{Complexity for encoding with} \ \bF_k \bA_{k,l} & = O
\left(\frac{M^2}{K} L \right),
\end{align*}
which shows that the block-wise encoding and decoding can be more computationally efficient as $K$ increases.

\begin{figure}[!t]\vspace{-1em}
\begin{center} \subfigure[Uncoded case with $J = 4$ workers]{\label{fig 2 ax}\includegraphics[width=0.45\textwidth]{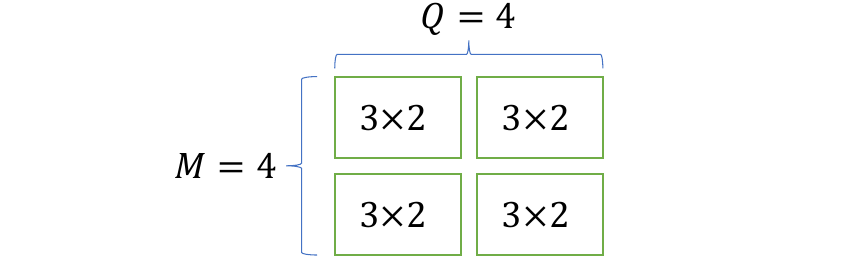}}
\subfigure[Coded case with $J = 6$ workers]{\label{fig 2 bx}\includegraphics[width=0.45\textwidth]{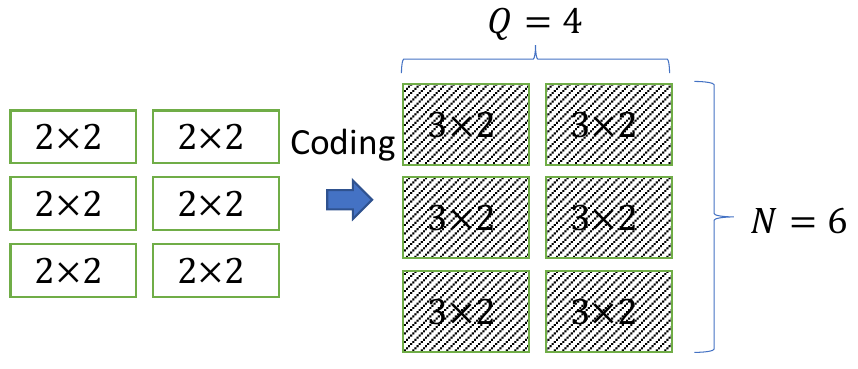}} \vspace{-1em}
\end{center}
\caption{An example of block-wise encoding and decoding. Note that shaded blocks represent coded data matrices, $\tilde \bA_{k,l}$.}\label{Fig:2}
\end{figure}

\begin{myexample}
To apply block-wise encoding and decoding with more workers, we need to properly divide tasks. Suppose that each worker can complete a task of 
$3 \times 2$ data matrix. If $M = 6$ and $Q = 4$, then there can be $J = 4$ workers with uncoded data matrices $\bA_{k,l} \in \uC^{3 \times 2}$ as shown in Fig.~\ref{Fig:2} (a). However, when coding is used, $\bA$ is now divided into 6 submatrices of size $2 \times 2$ as shown in Fig.~\ref{Fig:2} (b) (the left-hand side). Then, each submatrix is encoded into the corresponding coded submatrix, $\tilde \bA_{k,l} 
= \bF_k \bA_{k,l} \in \uC^{3 \times 2}$, and sent to the corresponding worker, which is one of $J = 6$ workers. 
\end{myexample}

In order to further reduce the complexity of encoding (or even without block-wise encoding/decoding), we can consider binary numbers for the elements of $\bF$ so that multiplication operators are not necessary, or a special case that the elements of $\bF$ are limited. In particular, with the elements of $\bF$ that are represented as powers of two, matrix multiplication can be replaced with a sequence of bit shifts and additions. This approach exploits the fact that multiplying a number by a power of two is equivalent to shifting its binary representation to the left. Thus, each multiplication operation can be replaced with a corresponding number of bit-shift operations, which can be performed more efficiently than multiplication.

In general, analog channel coding differs in certain aspects from digital channel coding. One key difference is that it does not require the message to be represented as a bit sequence. In the OTA-based approach, each worker can send a complex-valued vector, i.e., the outcome, $\tilde \bA_{k,l} \bx_l$, directly without any encoding to convert it into a bit sequence during the superposition phase. Another difference is the performance metric. Digital channel coding aims to minimize the probability of decoding errors, while analog channel coding in the OTA-based approach is to reduce the MSE. As a result, it is expected to have a performance in terms of MSE that is gradually degraded as $N$ approaches $M$.


\section{Performance Analysis}  \label{S:Anal}

In this section, we analyze the performance of the OTA-based approach under a few assumptions. A different setting from that in \cite{Reisizadeh19} \cite{Lee18} will be considered. In particular, thanks to the flexibility to adjust the computing load for each worker by dividing data matrix $\bA$ into multiple sub-matrices $\bA_{k,l}$ of different sizes, 
we assume that 
all the workers are able to complete their tasks within each round. Therefore, insufficient transmit power at workers is the only potential source of performance degradation in the OTA-based approach.

Key assumptions for the analysis in this section are as follows.
\begin{itemize}
\item[{\bf A1})] 
The $h_{k,l}$'s are independent and identically distributed (iid) and follow the Rayleigh distribution in the amplitude  so that
\be 
| h_{k,l}|^2 \sim \frac{1}{\sigma_h^2} e^{-\frac{| h_{k,l}|^2}{\sigma_h^2}},
\ee
i.e., $h_{k,l}$ is a CSCG random variable with $\uE[h_{k,l} ] = 0$ and $\uE[| h_{k,l}|^2 ] = \sigma_h^2$.

\item[{\bf A2})] 
Each element of $\by_{k,l} = \bA_{k,l} \bx_l \in \uC^{M_k \times 1}$ is an independent CSCG random variable with zero-mean and variance $\sigma_{k,l}^2$. Thus,
\be 
| [\by_{k,l}]_i|^2 \sim \frac{1}{\sigma_{k,l}^2} e^{-\frac{| [\by_{k,l}]_i|^2}{\sigma_{k,l}^2} } .
    \label{EQ:by}
\ee 
\end{itemize}

Note that when the additive white Gaussian noise (AWGN) channel is considered, the individual MSE is mainly dependent on $\psi_{k,l}$ and the performance analysis would be straightforward. Thus, we only consider the case of fading channels according to the assumption of {\bf A1} in this section. 
In addition, while it is possible to consider another distribution for 
$[\by_{k,l}]_i$, the assumption of {\bf A2} is utilized as an example for tractable analysis. This assumption remains reasonable, particularly when the elements of a large 
$\bA$ are iid, thanks to the central limit theorem.

\subsection{Uncoded Performance}



\begin{mylemma} \label{L:1}
Under the assumptions of {\bf A1} and {\bf A2}, for uncoded case, the probability that the individual MSE of worker $(k,l)$ is greater than $0$, which is called the individual outage probability, is given by 
\begin{align}
\epsilon_{k,l} = \Pr\left( \frac{\psi_{k,l}}{|h_{k,l}|^2} > \bar P_{k,l} \right)  
= \rho_{M_k} \left(\frac{\sigma_{k,l}^2}{\sigma_h^2 \bar P_{k,l}} \right),
    \label{EQ:L1}
\end{align}
where
\be 
\rho_n (s) = 1  - \prod_{i=1}^{n} \frac{i}{s+i}.
\ee
\end{mylemma}
\begin{IEEEproof}
It can be shown that 
\begin{align}
\epsilon_{k,l} & = 
\Pr\left( |h_{k,l}|^2 <\frac{\psi_{k,l}}{\bar P_{k,l} } \right) \cr 
& = 1 - \uE \left[ \exp \left( - \frac{\psi_{k,l}}{\sigma_h^2 \bar P_{k,l}}\right) \right] \cr 
& = 1 - \uE\left[ \exp \left( - \frac{\max_i | [\by_{k,l}]_i|^2 }{\sigma_h^2 \bar P_{k,l}}\right) \right] \cr 
& = 1 - \uE\left[ \exp \left( - s \max_{i \in \{1, \ldots, M_k\}} \nu_i \right) \right],
\end{align}
where $s = \frac{\sigma_{k,l}^2}{\sigma_h^2 \bar P_{k,l}}$ and $\nu_i$ is an independent exponential random variable with $\uE[\nu_i] = 1$. Using order statistics \cite{DavidBook}, we can show that 
\begin{align}
\Omega & = \uE\left[ \exp \left( - s \max_{i \in \{1, \ldots, M_k\}} \nu_i \right) \right] \cr 
& =  \int_0^\infty e^{-s x} M_k (1 - e^{-x} )^{M_k - 1} e^{-x} d x \cr 
& =  M_k  \int_0^\infty (1 - e^{-x} )^{M_k - 1} e^{-(s+1)x} d x \cr 
& = M_k \int_0^1 y^{M_k - 1} (1-y)^s d y \cr 
& = \frac{\Gamma (M_k ) \Gamma(s+1)}{ \Gamma (M_k + s+1) },
\end{align}
where the second last equality is due to the change of variables with $y = 1 -e^{-x}$ and $\Gamma (x) = \int_0^\infty t^{x-1} e^{-t} d t$ is the Gamma function.

Using the recurrence formula of Gamma function, i.e.,
$$
\Gamma(s) = \frac{\Gamma(s+n+1)}{s (s+1) \ldots (s+n)},
$$
and noting that $M_k$ is an integer, 
we have
\be 
\Omega = \frac{M_k!}{(s+1) \cdots (s+M_k)} = \prod_{i=1}^{M_k} \frac{i}{s+i},  
\ee 
which leads to \eqref{EQ:L1}.
\end{IEEEproof}

Using $\epsilon_{k,l}$, an upper-bound on the mean MSE can be found as
\begin{align}
\overline{\sMSE}& = \uE\left[\sum_k \sum_l \sMSE_{k,l} \right] 
+ \frac{N_0}{P_{\rm rx}} \sum_k M_k \cr 
& \le \sum_{k=1}^K \sum_{l=1}^L \uE[||\by_{k,l}||^2] \epsilon_{k,l}
+ \frac{N_0}{P_{\rm rx}} M,
    \label{EQ:UMMSE}
\end{align} 
where  the upper-bound is due to \eqref{EQ:bMSE0}. Thus, in order to have a lower mean MSE, it is crucial to have a low outage probability, $\epsilon_{k,l}$.

We can improve the bound in \eqref{EQ:UMMSE} as follows.

\begin{mytheorem}
Let
\be
\epsilon_{k,l} (z) = \rho_{M_k} \left(z \frac{\sigma_{k,l}^2}{\sigma_h^2 \bar P_{k,l}} \right), \ z \in [1/25, 1].
\ee 
Then, under the assumptions of {\bf A1} and {\bf A2}, we have
\begin{align}
\overline{\sMSE}
\le \sum_{k=1}^K \sum_{l=1}^L \uE[||\by_{k,l}||^2] 
\bar \epsilon_{k,l}+ \frac{N_0}{P_{\rm rx}} M,
    \label{EQ:UMMSE2}
\end{align} 
where 
\be
\bar \epsilon_{k,l} = 
\min_{z \in [1/25, 1]} 
\frac{1}{4} \left( \sqrt\frac{1}{z} - 1\right) 
+\frac{5 - \sqrt\frac{1}{z}}{4} \epsilon_{k,l} (z) .
    \label{EQ:bare}
\ee 
\end{mytheorem}
\begin{IEEEproof}
Let $\tau = \frac{\psi_{k,l}}{\bar P_{k,l}}$
and consider $z \in [1/25, 1]$.
According to \eqref{EQ:bMSE1} and Fig.~\ref{Fig:plt_MSE}, 
the normalized individual MSE is upper-bounded
by 1 for $|h_{k,l}|^2 < z \tau$ and $\frac{1}{4} 
\left( \sqrt\frac{1}{z} - 1 \right)$ for $|h_{k,l}|^2 \ge z \tau$. Thus,
we can show that
\begin{align}
\frac{\sMSE_{k,l}}{\uE[||\by_{k,l}||^2]} & \le
\Pr(|h_{k,l}|^2 < z \tau)  \cr 
& \quad + \frac{1}{4} 
\left( \sqrt\frac{1}{z} - 1 \right) (1-\Pr(|h_{k,l}|^2 < z \tau)). \qquad 
    \label{EQ:T2a}
\end{align}
As shown in Lemma~\ref{L:1}, we have
\be 
\Pr(|h_{k,l}|^2 < z \tau) = \epsilon_{k,l} (z).
    \label{EQ:T2b}
\ee 
Substituting \eqref{EQ:T2b} into \eqref{EQ:T2a}, the objective function in \eqref{EQ:bare} can be obtained. Then, by applying it to the first equation in \eqref{EQ:UMMSE}, we have \eqref{EQ:UMMSE2}. This completes the proof.
\end{IEEEproof}

Note that in \eqref{EQ:bare}, if $z = 1$, $\bar \epsilon_{k,l} (z)$ becomes $\epsilon_{k,l}$. Thus, $\bar \epsilon_{k,l}$ has to be less than or equal to $\epsilon_{k,l}$, meaning that \eqref{EQ:UMMSE2} is a tighter bound that \eqref{EQ:UMMSE}.

\subsection{Coded Performance}

To investigate the performance of coded systems, it is often useful to consider a randomized case. 
In this section, we will explore the notion of random coding, which provides a more comprehensive understanding of the behavior of analog channel coding for the OTA-based approach. In addition, to facilitate a performance comparison with the uncoded cases, we assume that $K$ and $L$ are fixed, while the number of rows in the sub-matrices, $M_k$, increases to $N_k$ (in this case, the number of workers remains unchanged, while more radio resources are required for analog coding).

For random coding, the following assumption will be considered.
\begin{itemize} 
\item[{\bf A3})] 
The elements of $\bF_k$ are assumed to be iid random variables with mean-zero and variance $\frac{1}{N_k}$.    
\end{itemize}

If $N_k$ and $M_k$ are sufficiently large, we can have
$$
\bF_k^\dagger \approx \bF_k^\rH,
$$
as $\bF_k^\rH \bF_k \approx \bI$. Thus, for simplicity, we only consider the asymptotic analysis with sufficiently large $N_k$ and $M_k$.

With analog coding, the outcome of worker $(k,l)$ is given by
\begin{align}
\tilde \by_{k,l} & = \tilde \bA_{k,l} \bx_l 
= \bF_k \bA_{k,l} \bx_l \cr  
& = \bF_k \by_{k,l} \in \uC^{N_k \times 1}.     
\end{align}
According to the assumption of {\bf A2}, we have
$\uE[\by_{k,l} \by_{k,l}^\rH] = \sigma_{k,l}^2 \bI$. From this, it can be shown that
\begin{align}
\uE[\tilde \by_{k,l} \tilde \by_{k,l}^\rH]
& = \uE[\bF_k \uE[\by_{k,l} \by_{k,l}^\rH \,|\, \bF_k] \bF_k^\rH] \cr 
& = \sigma_{k,l}^2 \uE[ \bF_k \bF_k^\rH] \cr 
& = \frac{M_k}{N_k} \sigma_{k,l}^2 \bI,
    \label{EQ:Cty}
\end{align}
where the second equality is due to the fact that $\bF_k$ and $\by_{k,l}$ are independent and the third equality
is due to the assumption of {\bf A3}.
Letting $r_k = \frac{M_k}{N_k}$, from \eqref{EQ:Cty},
the variance of of the elements of $\tilde \by_{k,l}$, denoted by $\tilde \sigma_{k,l}^2$, can be found as
\be 
\tilde \sigma_{k,l}^2 = r_k \sigma_{k,l}^2 \ (\le \sigma_{k,l}^2).
\ee 
For given $\bF_k$, each element of $\tilde \by_{k,l}$ is seen as a sum of CSCG random variables according to the assumption of {\bf A2}. Thus, 
we can assume that each squared element of $\tilde \by_{k,l}$ follows the distribution in \eqref{EQ:by} by replacing $\sigma_{k,l}^2$ with
$\tilde \sigma_{k,l}^2$.

\begin{mytheorem} 
Suppose that $M_k$ is fixed. Under the assumptions of {\bf A1}, {\bf A2}, and {\bf A3}, for a large $N_k$, we have
\be 
\epsilon_{k,l} = 1 - e^{- \beta (N_k)} ,
    \label{EQ:T3}
\ee 
where $\beta (N_k)$ is a decreasing function of $N_k$ and
\be
\beta (N_k) = O \left(\frac{\log N_k}{N_k} \right).
\ee 
\end{mytheorem}
\begin{IEEEproof}
From \eqref{EQ:L1}, by replacing $\sigma_{k,l}^2$ (for uncoded cases) with $\tilde \sigma_{k,l}^2$ (for coded cases), the outage probability can be obtained. 
Let 
$$
s = \frac{\tilde \sigma_{k,l}^2}{\sigma_h^2 \bar P_{k,l}}
= \frac{M_k}{N_k} \frac{\sigma_{k,l}^2}{\sigma_h^2 \bar P_{k,l}}.
$$
Then, for a fixed $M_k$, as $N_k$ increases, $s$ decreases. For $s< 1$, since $\frac{i}{s+i} \approx e^{-\frac{s}{i}}$, we have
\begin{align}
\epsilon_{k,l}
& = 1 - \prod_{i=1}^{N_k} \frac{i}{s+i} \cr 
& \approx 1 -  \exp \left( -s H_{N_k} \right) \cr 
& \approx 1 - \left( \frac{e^{-\bar \gamma}}{N_k} \right)^s,
    \label{EQ:a2}
\end{align}
where $H_n = \sum_{i=1}^{n}\frac{1}{i} \approx \ln n + \bar \gamma$ is a harmonic number and $\bar \gamma \approx 0.5772$ represents Euler's constant. The approximations in \eqref{EQ:a2} are tight as $N_k$ increases.

It can be shown that
\begin{align}
\log (1- \epsilon_{k,l} )& \approx - s (\log N_k + \bar \gamma) \cr 
& = - c_1 r_k \sigma_{k,l}^2 \left( \log N_k + \bar \gamma \right),
\end{align}
where $c_1= \frac{Q_l}{\sigma_h^2 \bar P_{k,l}}$, or
\begin{align}
\log (1- \epsilon_{k,l}) & \approx - c_1 \sigma_{k,l}^2 
\frac{M_k}{N_k} \left( \log N_k + \bar \gamma \right).
\end{align}
Letting
\be
\beta (N_k) = c_1 \sigma_{k,l}^2 
\frac{M_k}{N_k} \left( \log N_k + \bar \gamma \right) = O
\left( \frac{\log N_k}{N_k} \right),
\ee 
we can obtain \eqref{EQ:T3}.
\end{IEEEproof}

Furthermore, from \eqref{EQ:T3}, as $N_k$ increases, it can be readily shown that $\epsilon_{k,l}$ asymptotically becomes $\beta(N_k)$, i.e.,
\be 
\epsilon_{k,l} \to \beta (N_k) = O\left(\frac{\log N_k}{N_k} \right), 
    \label{EQ:ER_AC}
\ee
which eventually approaches 0 as $N_k \to \infty$. 
Since the MSE is the performance measure, from \eqref{EQ:UMMSE}, we have
\be 
\frac{\overline{\sMSE}}{\sum_k \sum_l \uE[||\by_{k,l}||^2]}
\to O\left(\frac{\log N_k}{N_k} \right),
    \label{EQ:AA}
\ee 
meaning that the MSE can gradually decrease at a rate of $\frac{\log N_k}{N_k}$.

\subsection{Comparison of OTA-based Approach and CM Schemes}

The OTA-based approach uses analog coding, while CM schemes use digital coding. Thus, the comparison becomes
challenging as they have different design principles, which  will be discussed in this subsection under the assumption  that the size of $\bA_{k,l}$ is the same for all workers and $\sigma_{k,l}^2 = \sigma^2$ for all $(k,l)$ in the assumption of {\bf A2}.

In Table~\ref{Tab:Comp}, we briefly compare two different types of approaches. When digital coding is employed, each complex-valued number has to be quantized and encoded. Based on the rate-distortion theory \cite{CoverBook}, under the assumption of {\bf A2}, we have
\be 
D = \sigma^2 2^{- R},
    \label{EQ:DR}
\ee 
where $D$ represents the distortion (in MSE) of each element of $\by_{k,l}$ and $R$ is the rate (or the number of bits per element). Thus, $\frac{D}{\sigma^2} = 2^{-R}$ becomes the normalized individual MSE due to quantization errors. 

\begin{table}[thb]    
    \centering
 \begin{tabular}{c||c|c}
         & OTA & CM Schemes  \\ \hline
Quantization error at workers & No & Yes \\
Workers' FEC       &  No & Yes \\ 
System-level coding & Yes & Yes \\   
Performance metric & MSE & Outage \\
\end{tabular}
\caption{Comparison of OTA-based approach with analog coding and CM scheme with digital coding.}
\label{Tab:Comp}
\end{table}

In CM schemes, workers typically employ a forward error correction (FEC) coding scheme to ensure reliable transmissions over noisy channels. However, FEC coding requires a bandwidth expansion due to the transmission of redundant bits, which can lead to increased radio resource requirements compared to the OTA-based approach. In contrast, the OTA-based approach does not utilize FEC coding, but instead incorporates the noise term into the total MSE, as demonstrated in \eqref{EQ:UMMSE}.

In order to compare with the OTA-based approach, in CM schemes, we assume that some workers become stragglers solely due to transmission errors (or deep fading) with 
the following outage probability that \eqref{EQ:hg} does not hold with $g_{k,l}
= \sqrt{\bar P_{k,l}}$:
\begin{align} 
P_{\rm out} & = \Pr \left( \eta > \log_2 \left(1 + \frac{|h_{k,l}|^2 \bar P_{k,l}}{N_0} 
\right) \right) \cr  
& = \Pr \left( |h_{k,l}|^2 < \frac{(2^\eta - 1) N_0}{\bar P_{k,l}} \right),
    \label{EQ:Pout}
\end{align}
where  $\eta$ is the transmission rate, which is proportional to $R$ in \eqref{EQ:DR}. The outage probability in \eqref{EQ:Pout} is the information outage probability under ideal FEC coding \cite{TseBook05}.
Let the average (receive) SNR be
$\sSNR = \frac{\sigma_h^2 \bar P_{k,l}}{N_0}$. Then, under the assumption of {\bf A1}, we have
\be 
P_{\rm out} = 1 - e^{- \frac{2^\eta - 1}{\sSNR}} \le 
\min\left\{\frac{2^\eta - 1}{\sSNR}, 1\right\}.
\ee
In \eqref{EQ:Pout}, the ideal achievable rate is assumed, while there are various overhead terms including redundant bits and performance loss due to finite-length codes to be taken into account. Thus, the relationship between $R$ and $\eta$ can be shown as
\be 
R = \eta_{\rm eff} = \alpha \eta, 
\label{EQ:Rae}
\ee 
where  $\eta_{\rm eff}$ is the effective transmission rate and $\alpha \in (0, 1)$ represents the efficiency\footnote{In \eqref{EQ:Rae}, it is assumed that each element of $\by_{k,l}$ is transmitted within one symbol duration for a fair comparision with the OTA-based approach.}
of coding systems.

When system-level coding is used with more workers (a total of $\bar J
\ (> J)$ workers) in CM schemes to mitigate stragglers, we can assume that the platform can decode if a sufficient number of the workers can successfully transmit without outage. Based on an optimistic assumption that the platform can build $\by$ if there are at least $J$ workers out of $\bar J$ without outage \cite{Mallick22},  the probability of decoding error can be given by
\begin{align} 
P_{\rm err} & = 1 - \sum_{n = J}^{\bar J} \binom{\bar J}{n} (1-P_{\rm out})^n P_{\rm out}^{\bar J-n} \cr 
& = \sum_{n=\bar J - J+1}^{\bar J}  \binom{\bar J}{n} P_{\rm out}^n  (1-P_{\rm out})^{\bar J-n}.
    \label{EQ:D1}
\end{align}
Note that
while the MSE serves as the performance metric for the OTA-based approach, the probability of decoding error is typically used as the performance indicator for CM schemes. Letting $\hat r = \frac{J}{\bar J}$, if $\hat r < 1 - P_{\rm err}$, an upper-bound on the probability of decoding error can be found as follows:
\begin{align} 
P_{\rm err} 
\le \exp \left(-\bar J~{\sf KL}(\hat r || 1-P_{\rm err} ) \right),
    \label{EQ:D2}
\end{align}
where ${\rm KL} (a||b) = a \log \frac{a}{b} + (1-a) \log \frac{1-a}{1-b}$ for $0 < a, b< 1$.

For a given $P_{\rm err}$, which is in general sufficiently small, we can find $P_{\rm out}$ from \eqref{EQ:D1} or \eqref{EQ:D2}, and then using \eqref{EQ:Rae}, we can obtain the MSE from \eqref{EQ:DR} and compare it with that of the OTA-based approach.




\section{Simulation Results}    \label{S:Sim}

In this section, we present the simulation results to see the performance in terms of normalized MSE (NMSE),
which is $\frac{\overline{\sMSE}}{\sum_k \sum_l \uE[||\by_{k,l}||^2]}$. All the simulations are carried out under the assumptions of {\bf A1} -- {\bf A3} with $\sigma_h^2 = 1$. It is also assumed that the elements of $\bA$ and $\bx$ are generated as independent CSCG random variable with zero mean and unit variance.

In addition, the size of $\bA_{k,l}$ is the same for all workers and the maximum transmit power of each worker is also the same, i.e., $\bar P_{k,l} = \bar P$. Thanks to the normalized channel gain, $\sigma_h^2 = 1$, we also have $P_{\rm rx} = \bar P$.
For convenience, the upper-bounds in \eqref{EQ:UMMSE} and \eqref{EQ:UMMSE2} are referred to as upper-bounds 1 and 2, respectively.

\begin{figure}[thb]
\begin{center}
\includegraphics[width=\figwidth]{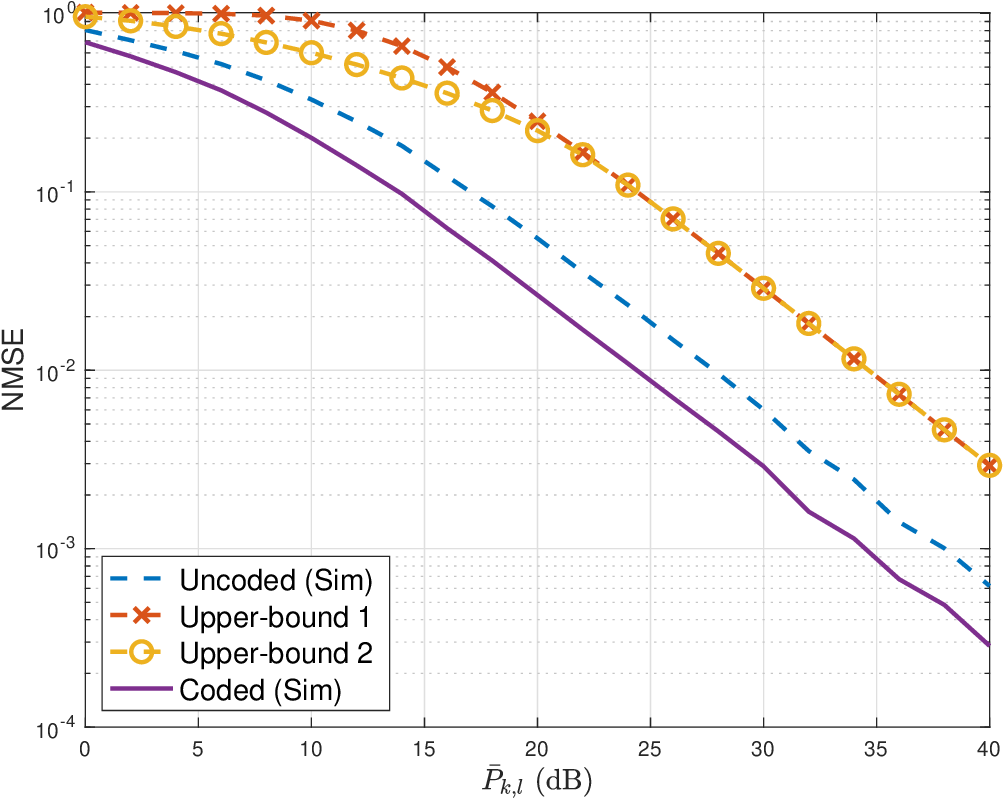} 
\end{center}
\caption{NMSE of the OTA-based approach as a function of the maximum transmit power when $K = L = 10$ and $M_k = Q_l = 10$.}
        \label{Fig:plt2}
\end{figure}

In Fig.~\ref{Fig:plt2}, the NMSE of the OTA-based approach is presented as a function of the maximum transmit power when $K = L = 10$ and $M_k = Q_l = 10$. For analog coding, $N_k = 3 M_k = 30$ is considered, while each element of $\bF$ is an independent CSCG random variable.
Clearly, it is shown that the NMSE decreases with the transmit power and analog coding can help reduce the NMSE. 
Upper-bound 2 is tighter than upper-bound 1 when the maximum transmit power is low, while they are close to each other when the maximum transmit power increases.

\begin{figure}[thb]
\begin{center}
\includegraphics[width=\figwidth]{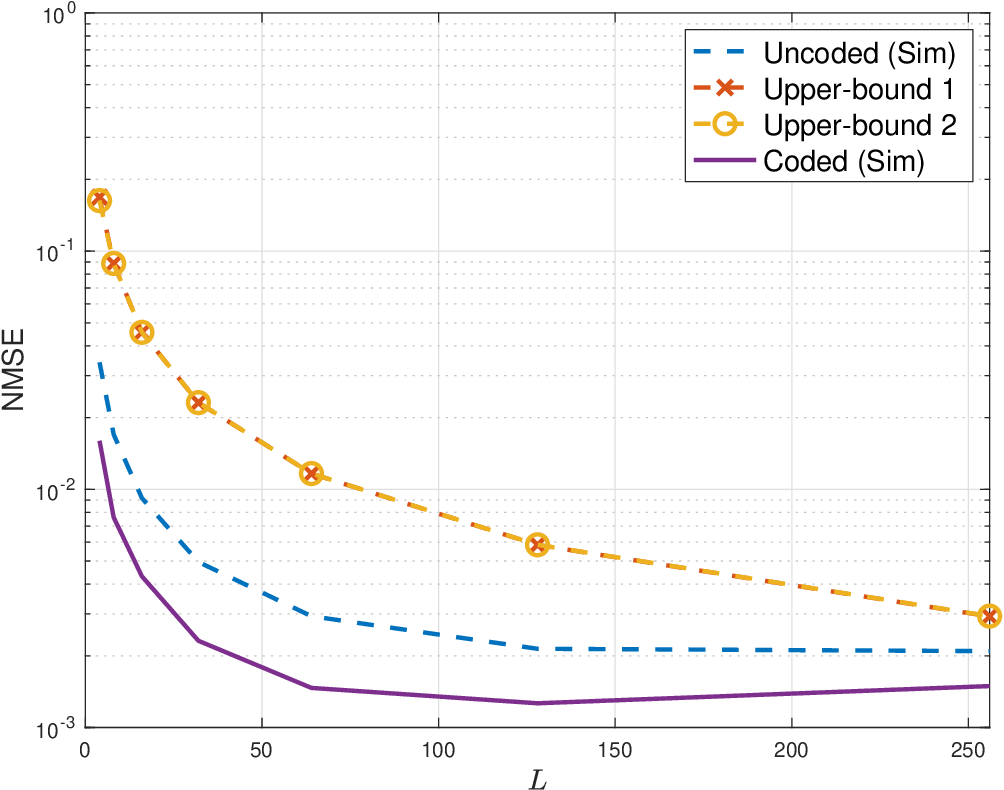} 
\end{center}
\caption{NMSE of the OTA-based approach as a function of $L$ when $\sSNR = 30$ dB, $K = M_k = 10$
and $L \in \{4, 8, \ldots, 256\}$, while the number of columns of $\bA$, $Q_l L = 2^8 = 256$, is fixed.}
        \label{Fig:plt5}
\end{figure}

Column-wise partitioning plays a crucial role
in reducing the computing load of workers and  increasing the number of workers  without requiring more radio resources in the OTA-based approach. Thus, it is interesting to see the impact of increasing the number of workers in each group, i.e., $L$, on the performance, which is shown in Fig.~\ref{Fig:plt5}, where $\sSNR = 30$ dB, $K = M_k = 10$,
and $Q_l \in \{4, 8, \ldots, 256\}$, while $Q_l L = 2^8 = 256$ is fixed. Thus, the number of columns of $\bA_{k,l}$ decreases with $L$, while the size of $\bA$ is fixed.
It can be observed that the NMSE decreases
with $L$. Clearly, although there are more workers, thanks to OTA computation, no additional radio resource is required in transmitting workers' outcomes to the platform and a better performance is achieved. We can also see that analog coding with $N_k = 3 M_k = 30$ can further reduce the NMSE.

\begin{figure}[thb]
\begin{center}
\includegraphics[width=\figwidth]{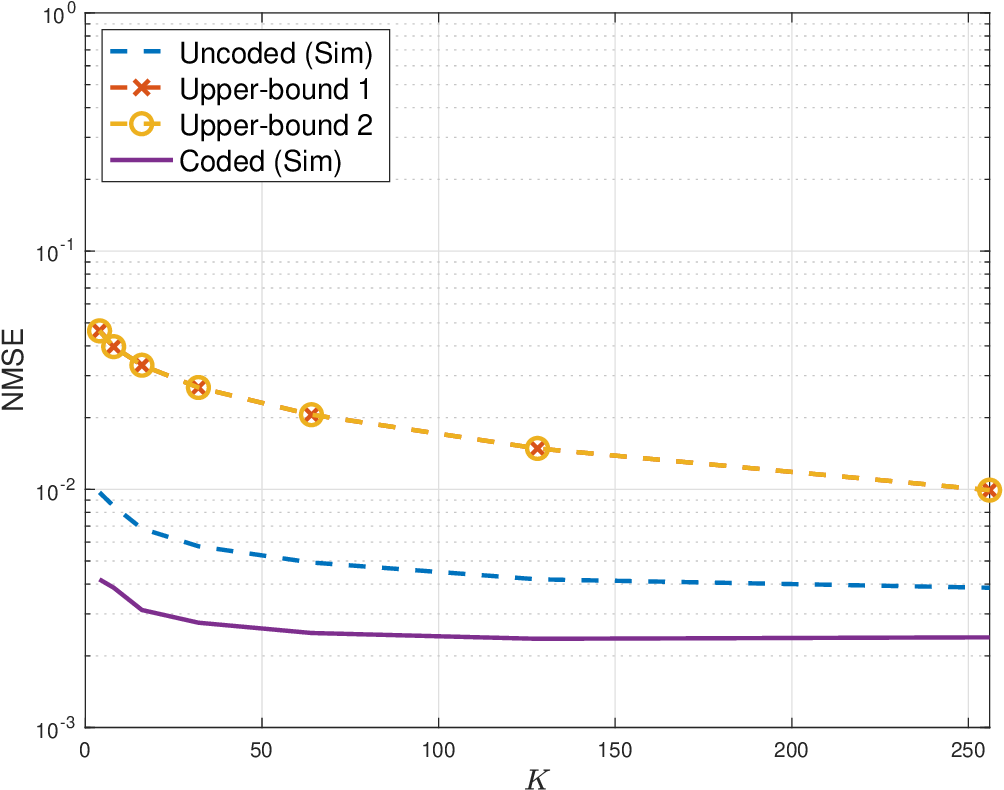} 
\end{center}
\caption{NMSE of the OTA-based approach as a function of $K$ when $\sSNR = 30$ dB,
$L = Q = 10$,
and $K \in \{4, 8, \ldots, 256\}$, while the number of rows of $\bA$, $K M_k = 2^8 = 256$, is fixed.}
        \label{Fig:plt3}
\end{figure}

For a given size of $\bA$, we can also increase the number of workers by increasing $K$ while the number of rows, $K M_k$, remains unchanged. To this end, we consider the case that $ K M_k = 256$, while $K \in \{4, 8, \ldots, 256\}$ when  $L = Q = 10$ and present the NMSE result in Fig.~\ref{Fig:plt3}. It is also shown that the NMSE decreases with $K$ or more workers, while the size of data matrix $\bA$ is fixed.

Note that increasing the number of column partitions, $L$, has a more significant impact on decreasing the NMSE compared to increasing the number of row partitions, $K$, as evidenced in the comparison between Figs.~\ref{Fig:plt5} and \ref{Fig:plt3}. As $L$ increases, the number of columns in each worker, $Q_l$, decreases, meaning that each element of $\by_{k,l} = \bA_{k,l} \bx_l$ decreases in amplitude, leading to a lower outage probability and potential improvement in NMSE. 
When $K$ increases, the number of elements in $\by_{k,l}$, $M_k$, decreases, which can potentially reduce $\max_i |[\by_{k,l}]_i|^2$ and lower the outage probability. However, the resulting performance improvement is limited. Therefore, it is more effective to increase the number of column partitions, $L$, when the total number of workers, $J = KL$, is fixed.

In Figs.~\ref{Fig:plt5} and \ref{Fig:plt3}, we have shown that the performance of the OTA-based approach can be improved by increasing more workers. A further performance improvement can be achieved using analog coding without increasing the number of workers.

\begin{figure}[!t]\vspace{-1em}
\begin{center} \subfigure[$\sSNR = 20$ dB]{\label{fig 8 ax}\includegraphics[width=0.45\textwidth]{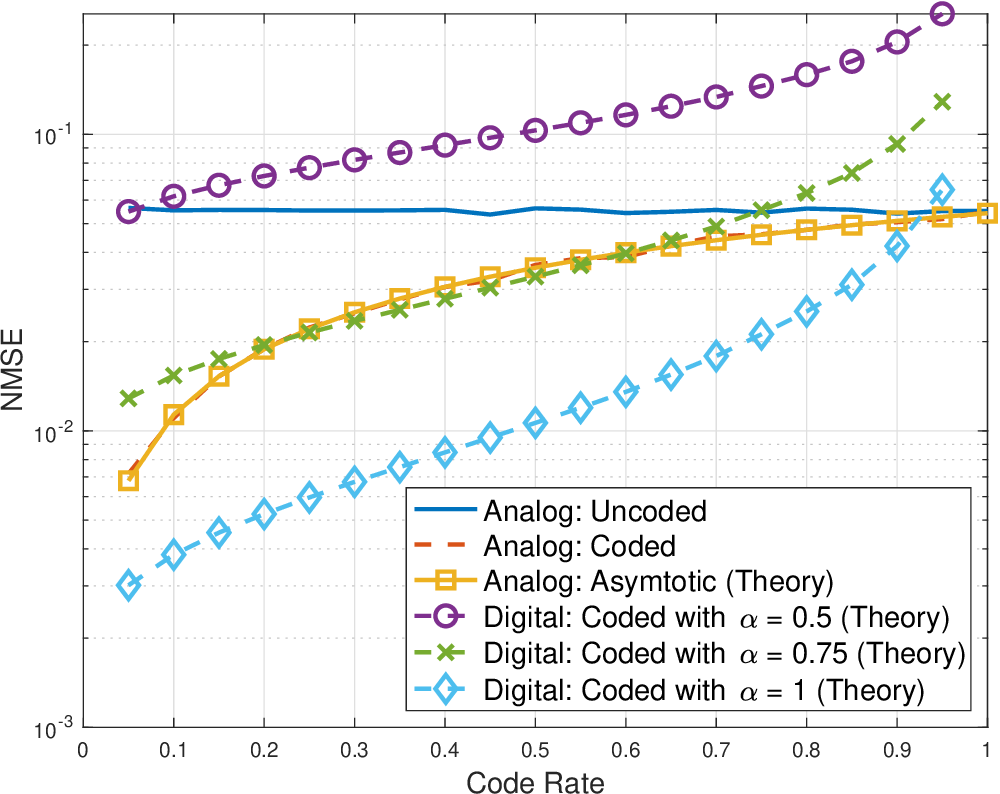}}
\subfigure[$\sSNR = 30$ dB]{\label{fig 8 bx}\includegraphics[width=0.45\textwidth]{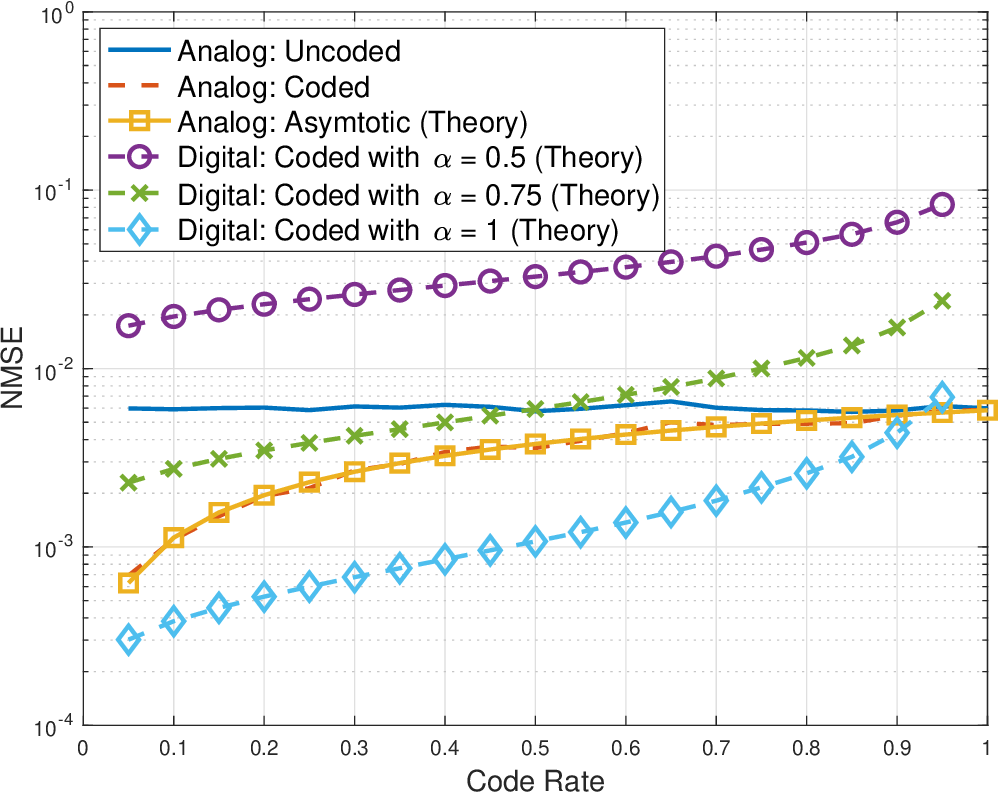}} \vspace{-1em}
\end{center}
\caption{NMSE as a function of code rate $r = \frac{M_k}{N_k}$ when $K = L = M_k = Q_l = 10$: (a) $\sSNR = 20$ dB; (b) $\sSNR = 30$ dB.}
    \label{Fig:plt1}
\end{figure}

For comparison, the NMSE of CM schemes (i.e., digital communication techniques) is also shown in Fig.~\ref{Fig:plt1},
with $K = L = M_k = Q_l = 10$, where the NMSE as a function of code rate $r = \frac{M_k}{N_k}$. It is shown that the NMSE decreases with decreasing $r$ and the asymptotic analysis in \eqref{EQ:AA} agrees with the simulation results.

With $\sSNR = 20$ dB, Fig.~\ref{Fig:plt1} (a) shows that CM schemes with $\alpha = 1$ can provide a lower NMSE than the OTA-based approach with analog coding. However, with $\alpha = 0.75$,  the performances of the OTA-based approach and CM schemes are comparable, while the performance of CM schemes becomes poor when $\alpha = 0.5$. With $\sSNR = 30$ dB, as shown in Fig.~\ref{Fig:plt1} (b), the performance of CM schemes with $\alpha = 0.75$ becomes worse than that of the OTA-based approach.

Note that in Fig.~\ref{Fig:plt1}, we do not account for the increase in radio resources in CM schemes when column-wise partitioning is used. Since each worker requires a dedicated channel, column-wise partitioning results in a higher allocation of radio resources, leading to a decrease in $\alpha$ in \eqref{EQ:Rae} and, consequently, a worse performance in NMSE.

However, the proposed approach can only achieve a reasonably low NMSE with a high SNR (between 20 and 30 dB). For instance, when the SNR is not sufficiently high, the resulting NMSE deteriorates, as shown in Fig.~\ref{Fig:plt2}, indicating that accurate computational results are not anticipated. Therefore, in a low SNR regime, it is not recommended to use the proposed approach. Instead, digital communication techniques should be considered, despite the associated high bandwidth expansion as discussed above.

\section{Concluding Remarks}    \label{S:Con}


In this paper, we studied the application of OTA computation to distributed matrix-vector multiplications in DML. Thanks to OTA computation, column-wise partitioning can effectively adjust the computing load to workers of different computing capabilities while simultaneously increasing the number of workers without requiring additional radio resources. This approach offers significant advantages in terms of computational granularity, allowing each worker to handle tasks within their specific capabilities. In addition, our analysis of the MSE provided valuable insights into the performance characteristics of the OTA-based approach, considering factors such as maximum transmit power and number of workers. Furthermore, we proposed and evaluated an analog coding approach tailored for OTA-based computation, demonstrating its effectiveness in improving the performance in terms of MSE. 

There are several intriguing research topics to explore in the future. One such topic is load balancing incorporating the known statistical properties of wireless channels, which can provide valuable insights into OTA-based approaches with coding. Another interesting area is the application of the OTA-based approach to distributed sparse matrix-vector multiplications. Sparse matrices offer unique advantages, such as low transmit power requirements, which can effectively reduce the MSE even with limited maximum transmit power for distributed workers.

\bibliographystyle{ieeetr}
\bibliography{dml}
\end{document}